\documentclass[]{article}

\usepackage{graphicx}
\usepackage{wrapfig}
\usepackage{authblk}
\usepackage{amsmath}
\usepackage[margin=1in]{geometry}

\begin{document}

\title{Low Frequency Tilt Seismology with a Precision Ground Rotation Sensor}

\author[1]{M.P.Ross}
\author[1]{K.Venkateswara}
\author[1]{C.A. Hagedorn}
\author[1]{J.H. Gundlach}
\author[2]{ J.S. Kissel}
\author[2]{J. Warner}
\author[2]{H. Radkins}
\author[2]{T.J. Shaffer}
\author[3]{M.W. Coughlin}
\author[1]{P. Bodin}

\affil[1]{University of Washington, Seattle, WA}
\affil[2]{LIGO Hanford Observatory, Richland, WA}
\affil[3]{Harvard University, Cambridge, MA}
\maketitle
\begin{abstract}
\quad We describe measurements of the rotational component of teleseismic surface waves using an inertial high-precision ground-rotation-sensor installed at the LIGO Hanford Observatory (LHO). The sensor has a noise floor of 0.4\,nrad$/ \sqrt{\rm Hz}$  at 50\,mHz and a translational coupling of less than 1\,$\mu$rad/m enabling translation-free measurement of small rotations. We present observations of the rotational motion from Rayleigh waves of six teleseismic events from varied locations and with magnitudes ranging from M6.7 to M7.9. These events were used to estimate phase dispersion curves which shows agreement with a similar analysis done with an array of three STS-2 seismometers also located at LHO.
\end{abstract}

\section{Introduction}

\quad Measurements of the rotational components of seismic motion have gained increasing attention due to their usefulness in seismology \cite{Lee01052009} along with their applications in the seismic isolation systems of ground based gravitational-wave detectors \cite{lantz}. One geophysical application of these measurements is the ability to fully characterize surface waves, specifically their phase velocities, using devices placed at a single station, in contrast to the current practice of utilizing arrays of translational seismometers. These point-like measurements can lead to useful local seismological information that may be impractical to obtain using arrays while also simplifying measurement schemes \cite{seismic}. Rotational measurements also promise to improve seismic source-inversion problems \cite{se-7-1467-2016}. Additionally, rotation sensors of sufficient sensitivity can be used to reduce the tilt-noise in horizontal seismometers \cite{venk2016}, leading to improved active seismic isolation in gravitational-wave detectors.

The rotational components are henceforth referred to as the two `tilts' about two orthogonal horizontal axes and a `torsion' about a vertical axis. The tilt component along the radial direction of a teleseismic wave at a distant site can be estimated as  $\kappa A =2\pi A/\lambda$, where $A$ is the amplitude of the vertical displacement, $\kappa$ is the seismic wavenumber, and $\lambda$ is the seismic wavelength \cite{Lee01052009}. For amplitudes of 30\,$\mu$m and wavelengths of $\sim 70$\,km (typical for M7 earthquake halfway around the world, measured at Hanford, WA, USA), this corresponds to a tilt amplitude of 3\,nrad, making this measurement very challenging. Measurements of the rotational component of teleseismic events have been recorded previously for large earthquakes using sensitive ring-laser gyroscopes \cite{GRL:GRL13842, Belfi2012} in the torsional direction (about the vertical axis) or in the near-field regions using strong-motion sensors \cite{GRL:GRL28234}. These measurements have had low signal to noise ratios or require strong local events to observe the rotational signals. The low sensitivity of the rotational output to other degrees of motion, particularly horizontal motion, is critical as it can mimic rotational signals leading to incorrect wave parameter estimates. In the above example of teleseismic waves, a displacement rejection much better than $\kappa$ ($\sim 10^{-4}$\,rad/m) is required in order to accurately measure the tilt component.

We present the observations of the rotational component about a horizontal axis of teleseismic events with magnitudes ranging from M6.7 to M7.9 originating from different parts of the world. The rotation-sensor is described in detail in \cite{venk2014}. The measurements have high signal-to-noise ratio (SNR) and have negligibly small coupling from horizontal motion. In the next section, we describe the sensors used and their location. Subsequently, we present the framework used to characterize the surface waves. We then discuss rotational and translational data recorded from the examined earthquakes, focusing on a M7.9 in Papua New Guinea. The data from the rotation-sensor and an array of STS-2 seismometers at the LIGO Hanford Observatory are then analyzed to infer the angle of arrival and the Rayleigh-wave velocity at the site with high precision.

\section{Instrumentation}
\begin{wrapfigure}{}{.5\textwidth}
\includegraphics[width=.5\textwidth]{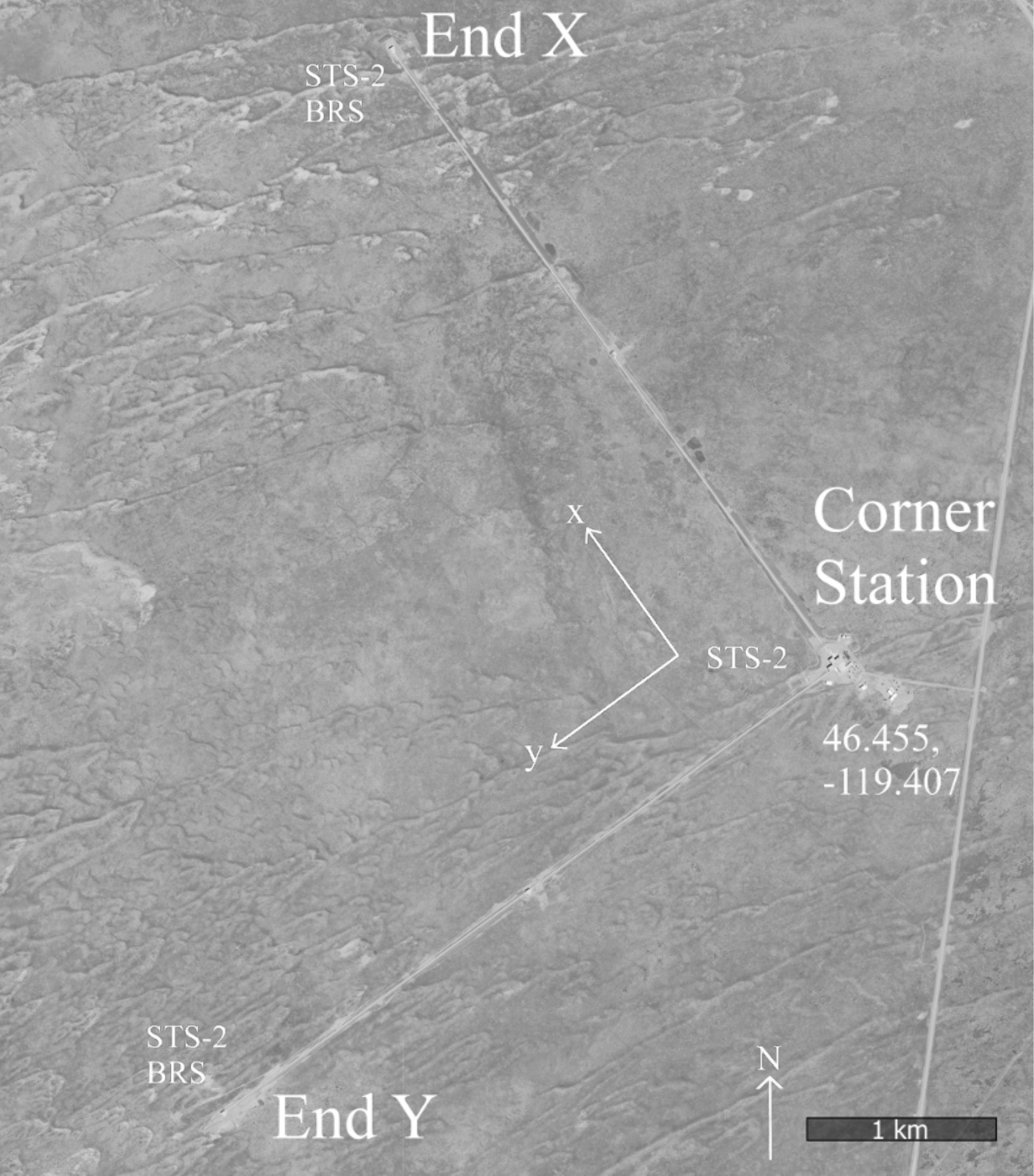}
\caption{Satellite image of LIGO Hanford Observatory showing the two arms, Arm-X running in the northwest direction, Arm-Y running in the southwest direction. Satellite image courtesy of USGS.}
\label{Map}
\end{wrapfigure}

\quad A set of three seismometers and two ground rotation sensors are operated as part of the Advanced LIGO seismic isolation system at the LIGO Hanford Observatory (LHO) \cite{ligo1,ligo2}. The observatory is chiefly comprised of three buildings: a Central building (called the Corner Station) and two End Station buildings located 4-km away. It forms a large L--shape with one arm (EndX) running in the northwest direction and the other (EndY) orthogonally oriented along the southwest direction as shown in Fig. \ref{Map}.

The seismometers used in this study were three-component STS-2 seismometers, located one each at the corner station and at the X and Y End Stations. Along with these, two single--axis ground rotation sensors (referred to as BRS-Beam Rotation Sensor) were installed at LHO, one at each End-Station, measuring rotation about the axis orthogonal to their respective arm axes. These are located approximately 1\,m from the seismometers. Following the coordinate convention at LHO, we assign the X-axis to be parallel to the X-arm, Y-axis to be parallel to the Y-arm and Z-axis to be aligned with the gravitational vertical. Tilts measured along the Y-axis (or about the X-axis) are referred to as Y-tilt or angle. 

The BRS used in this study consists of a 1\,m long beam-balance suspended from 15\,$\mu$m thick flexures with a fundamental resonance frequency of 7.7\,mHz. The angle of the beam relative to the ground is measured with a high-sensitivity autocollimator whose noise floor is $\sim$0.2\,nrad/$\sqrt{\rm Hz}$ above 60\,mHz. For more details on the BRS see \cite{venk2014}. As the ground tilts at frequencies above the resonance of the beam balance, the beam stays inertial, hence the autocollimator measures the ground tilt. 

Translational acceleration coupling of a BRS is minimized by adjusting the center of mass to be close to the pivot point of the flexures. At frequencies greater than the resonance frequency of the beam-balance, the strength of this coupling can be approximated with $M\delta /I$ where M is the mass of the balance, I is the moment of inertia, and $\delta$ is the distance from center of mass to the pivot point \cite{venk2014}. For the BRS at EndX, $\delta$ was found to be 30\,$\mu$m, after installation. Schedule constraints did not permit further adjustments. This $\delta$ leads to a translation coupling of $2 \times 10^{-4}$\,rad/m, which pollutes the tilt signal with translation at the frequencies of interest for teleseismic events. For the BRS at EndY, $\delta$ was tuned to less than 0.5\,$\mu$m during installation. The very small value of $\delta$ for this instrument leads to a translation coupling of less than $1 \times 10^{-6}$\,rad/m and allows for precise measurement of the tilt component of low frequency seismic waves.

\begin {figure}[!h]
\centering
\includegraphics[width=.85\textwidth]{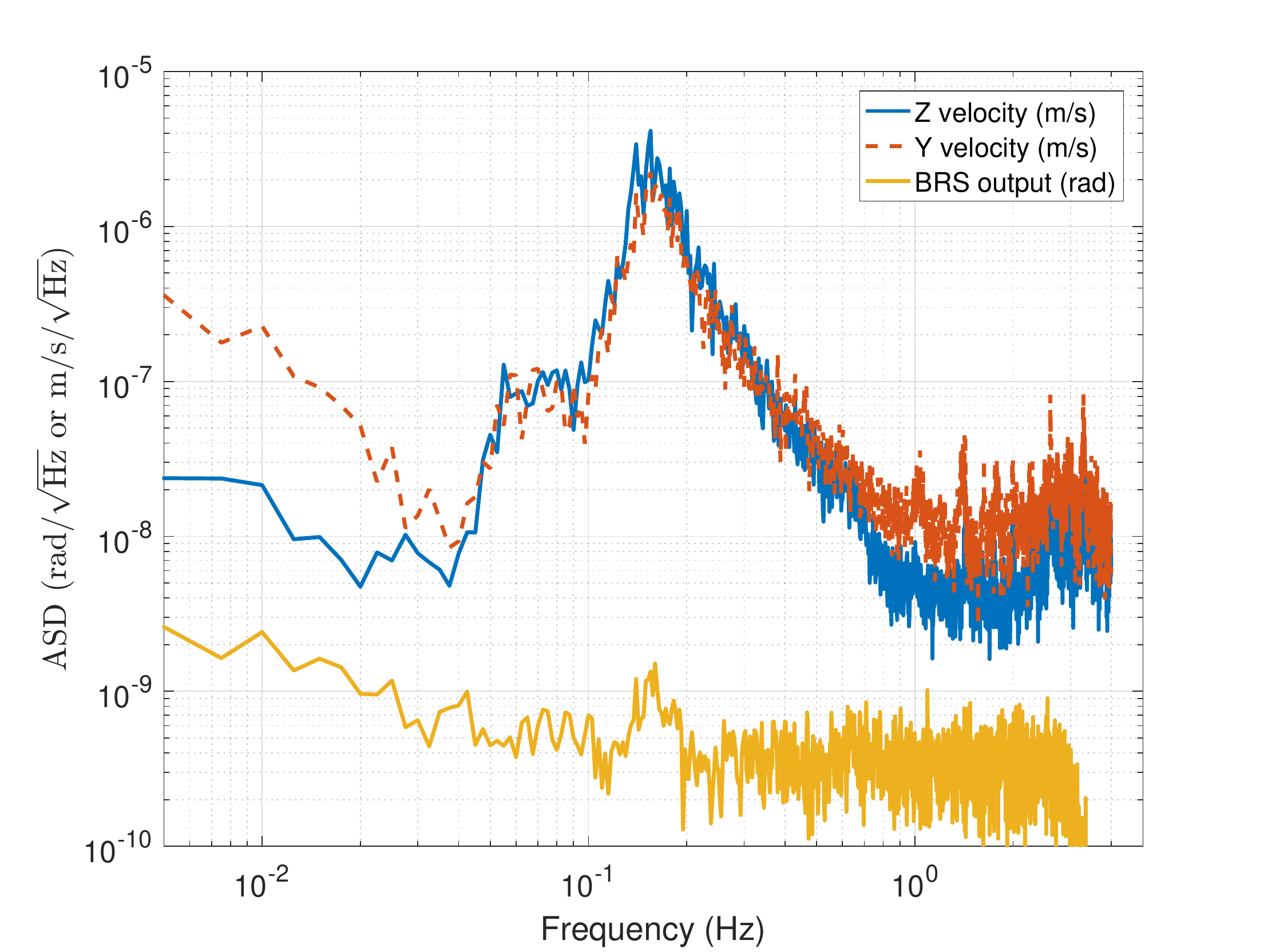}
\caption{Amplitude spectral density of the signals during a 2000\,s stretch before the arrival of the signals from the M7.9 earthquake on Dec. 17, 2016.}
\label{PNGASDQuiet}
\end{figure}
\pagebreak
Fig. \ref{PNGASDQuiet} shows the Amplitude Spectral Density (ASD) of the BRS signal along with two axes of the collocated seismometer before arrival of the Papua New Guinea earthquake on Dec. 17, 2016, recorded at 10:30:00 UTC for 2000 s. Wind-speeds were below 3\,m/s during this period and therefore the BRS curve is representative of the noise floor of the instrument, which can resolve tilt amplitudes as small as 0.1\,nrad for periods in the range of 10-50 seconds.

In this study, we present phase velocity estimations derived from tilt component measurements with the BRS at EndY of multiple teleseismic waves with a variety of angles of incidence and distances of origin from the facility. We did not analyze the data from the BRS at EndX due to its large horizontal motion coupling. In principle, if the value of $\delta$ were known precisely enough, this coupling could be removed from the data using the horizontal seismometer channel. The current uncertainty on $\delta$ introduces large errors in the rotation measurement with BRS-X, thus it is not used in this study. For the remainder of this paper the term BRS refers to only the BRS at EndY.
\section{Theoretical Framework}
\quad The plane wave solution of a Rayleigh wave has three translational and three rotational components denoted by $(u_x,u_y,u_z,\theta_x,\theta_y,\theta_z)$ respectively where subscripts denote the axes translations are along and rotations are about. They can be described by:
\begin{align}
u_x({\bf r} ,t)&=\alpha sin(\zeta) cos(\psi) cos(\omega t-{\bf k}\cdot{\bf r} + \phi)\\
u_y({\bf r} ,t)&=\alpha sin(\zeta) sin(\psi) cos(\omega t-{\bf k}\cdot{\bf r} + \phi)\\
u_z({\bf r} ,t)&=\alpha cos(\zeta) cos(\omega t-{\bf k}\cdot{\bf r}+\pi/2 + \phi)\label{ZEq}\\
\notag \\
\theta_x({\bf r} ,t)&=\frac{\partial u_z}{\partial y}=\alpha \kappa cos(\zeta) sin(\psi) cos(\omega t-{\bf k}\cdot{\bf r} + \phi)\label{TiltXEq}\\
\theta_y({\bf r} ,t)&=-\frac{\partial u_z}{\partial x}=-\alpha \kappa cos(\zeta) cos(\psi) cos(\omega t-{\bf k}\cdot{\bf r} + \phi)\\
\theta_z({\bf r} ,t)&=\frac{1}{2}\Big(\frac{\partial u_y}{\partial x}-\frac{\partial u_x}{\partial y}\Big)=0
\end{align}

Where $\alpha$ is the amplitude of the wave, $\zeta$ is the ellipticity angle, $\psi$ is the angle of propagation, $\omega$ is the angular frequency of the wave, $\kappa$ is the wavenumber, $\phi$ is the phase of the wave, {\bf r} is the position, and {\bf k} is the wave vector $\kappa (cos(\psi),sin(\psi),0)$ \cite{seismic}. With the assumption that the wavenumber is constant over the region of interest, this gives us six parameters to fully describe the surface field of a Rayleigh wave. Measuring three axes of translation at a single point alone cannot fully characterize the wave, whereas measuring the three translations and two tilts is sufficient. Measurement of the $\theta_z$ is useful to distinguish between Love waves and Rayleigh waves but is more difficult since Love waves typically have even smaller amplitudes. 

If the relevant components of a Rayleigh wave can be measured and the plane-wave assumption holds, then the phase velocity of the wave can be computed by:
\begin{equation}
v=-\frac{\dot{u}_z}{\theta_x} sin(\psi) \label{BRS V}
\end{equation}
Where $v$ is the wave velocity, $\dot{u}_z$ is the magnitude of the vertical seismometer velocity, $\theta_x$ is the tilt about the x--axis, and $\psi$ is the direction of propagation with respect to the x--axis.

The phase velocity and angle can also be computed from the phase difference between a network of three seismometers located in an L-shape, as shown in Fig. 1, by:
\begin{equation}
v=\frac{d}{\sqrt{\Delta t_x^2+\Delta t_y^2}} \label{SEI V}
\end{equation}
\begin{equation}
\psi=arctan(\frac{\Delta t_y}{\Delta t_x}) \label{SEI A}
\end{equation}
\lowercase{w}here $d$ is the distance between the corner seismometer and the end seismometers, $\Delta t_x$ and $\Delta t_y$ are the delays of the time of arrival between the seismometers in the x--direction and y--direction, respectively.

\section{Earthquake signals}
\quad A selection of recent earthquakes were chosen for this study, which showed prominent teleseismic tilt signals. Earthquake magnitude, distance from measurement site, activity (both environmental and anthropogenic) at the site, and angle of incidence can all influence whether or not the signal is prominent. Due to the geometry of the array and rotation sensor, our measurements were insensitive to waves traveling along EndX (NW-SW). Additionally, our instruments are sensitive to tilts driven by wind and human activity which eliminated many candidate events in our search.

\begin {figure}[h]
\centering
\includegraphics[width=\textwidth]{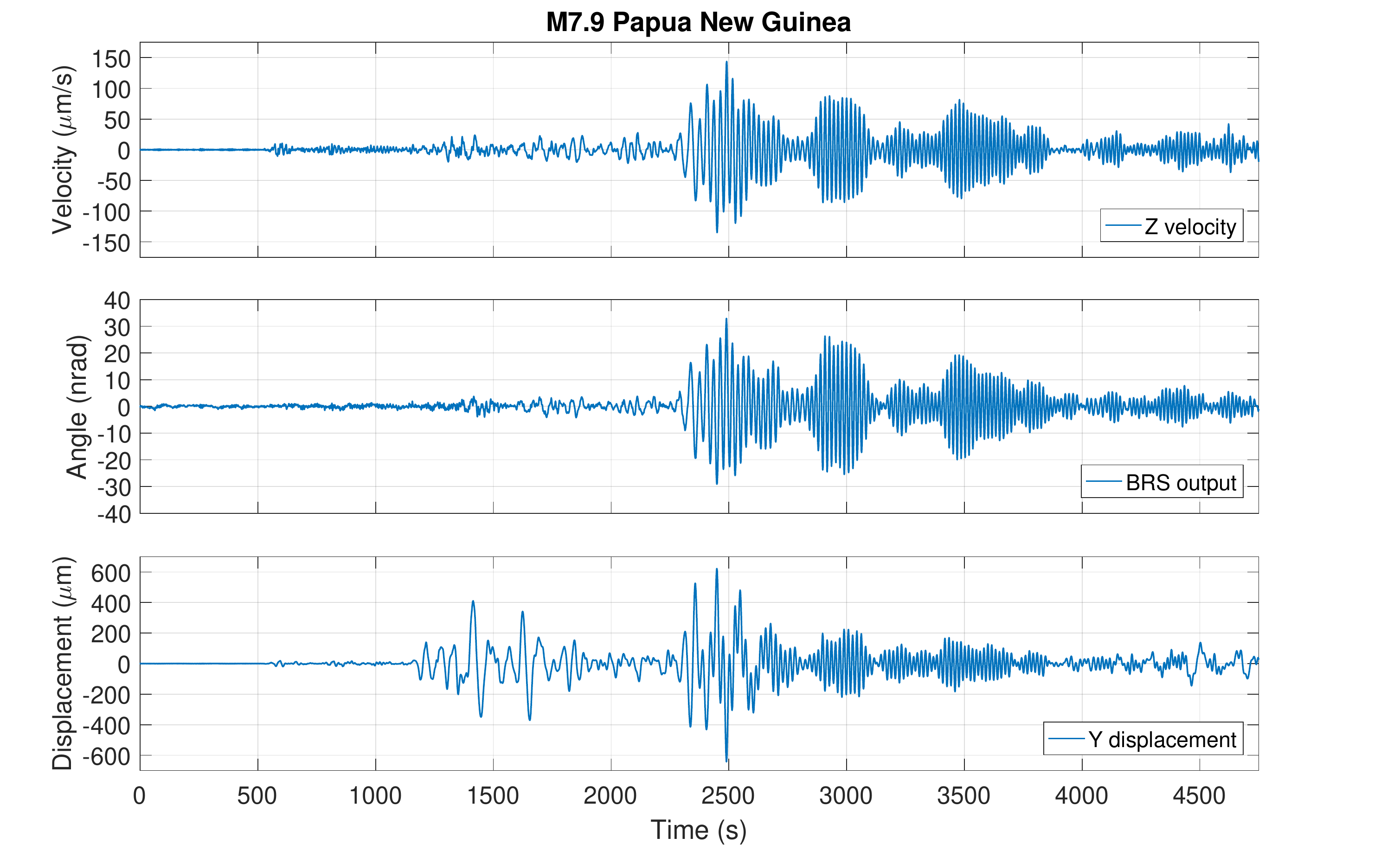}
\caption{Time series of the December 17, 2016 magnitude 7.9 earthquake east of New Ireland, Papua New Guinea, band-pass filtered between 10\,mHz and 300\,mHz. The x-axis starts at the earthquake origin time of 10:51:12 UTC.}
\label{PNGTS}
\end{figure}

\begin {figure}[!h]
\centering
\includegraphics[width=.65\textwidth]{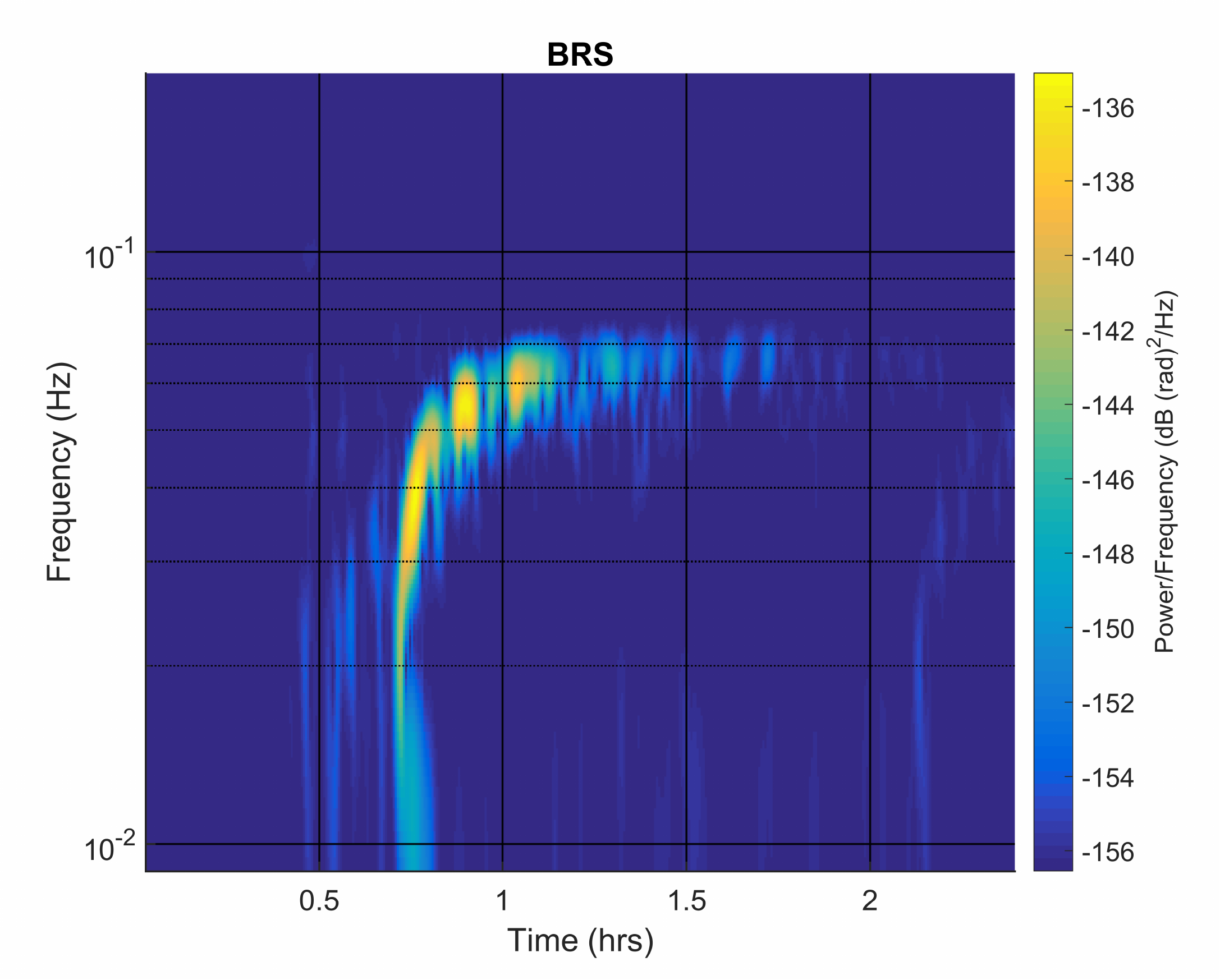}
\caption{Spectrogram of rotational displacement of the M7.9 Papua New Guinea earthquake on December 17, 2016. The x-axis at the earthquake origin time of 10:51:12 UTC. }
\label{BRSspec}
\end{figure}

\begin {figure}[!h]
\centering
\includegraphics[width=.65\textwidth]{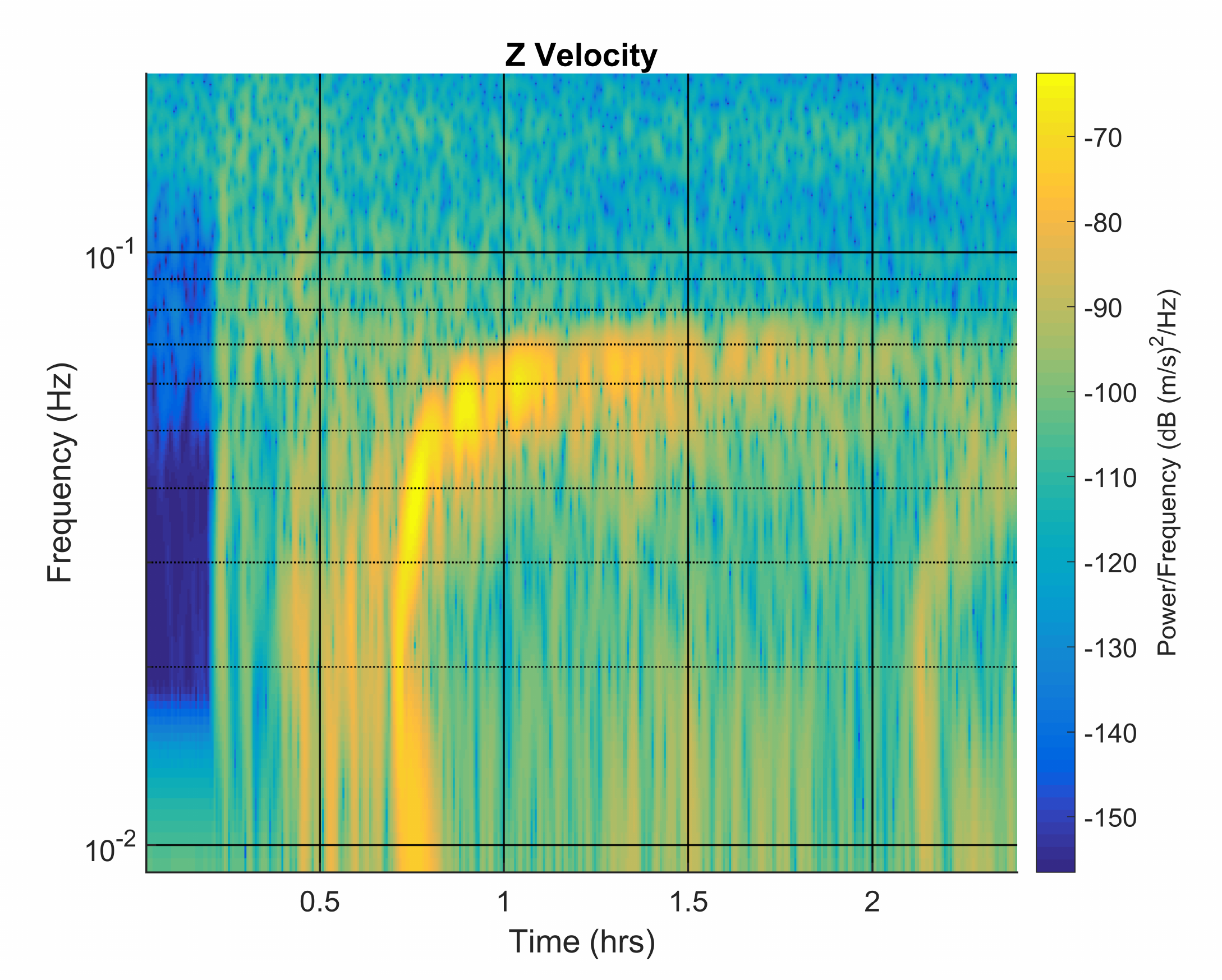}
\caption{Spectrogram of vertical displacement of the M7.9 Papua New Guinea earthquake on December 17, 2016. The x-axis starts 1300\,s before the earthquake origin time of 10:51:12 UTC.}
\label{Zspec}
\end{figure}

Fig. \ref{PNGTS} shows the time trace of the STS-2 seismometer and the BRS at the EndY from the December 17, 2016 magnitude 7.9 earthquake east of New Ireland, Papua New Guinea. The signals are band-passed between 10\,mHz and 300\,mHz by a 4th order Butterworth filter to focus on the earthquake signal. The first plot is the velocity along the Z-direction, and the second plot is the angle about the X-direction measured by BRS. The signals look nearly identical, as one expects from Eqs. \ref{ZEq}, \ref{TiltXEq} if Rayleigh waves dominate the signal. The third plot shows the displacement measured along the Y-direction. Note that the Y displacement is qualitatively different from the signal in the BRS. Around t = 1200 seconds, a large amplitude signal is visible in the Y-displacement, with much smaller corresponding signals in X-angle or Z-velocity, suggesting a shear wave. The small signal in BRS shows a displacement rejection of better than $\sim 3\times10^{-6}$\,rad/m which translates to less than 3 percent error introduced from translation coupling. Figures \ref{BRSspec} and \ref{Zspec} show spectrograms of the two instruments starting at the earthquake origin time of 10:51:12 UTC. 

\begin {figure}[!h]
\centering
\includegraphics[width=1.1\textwidth]{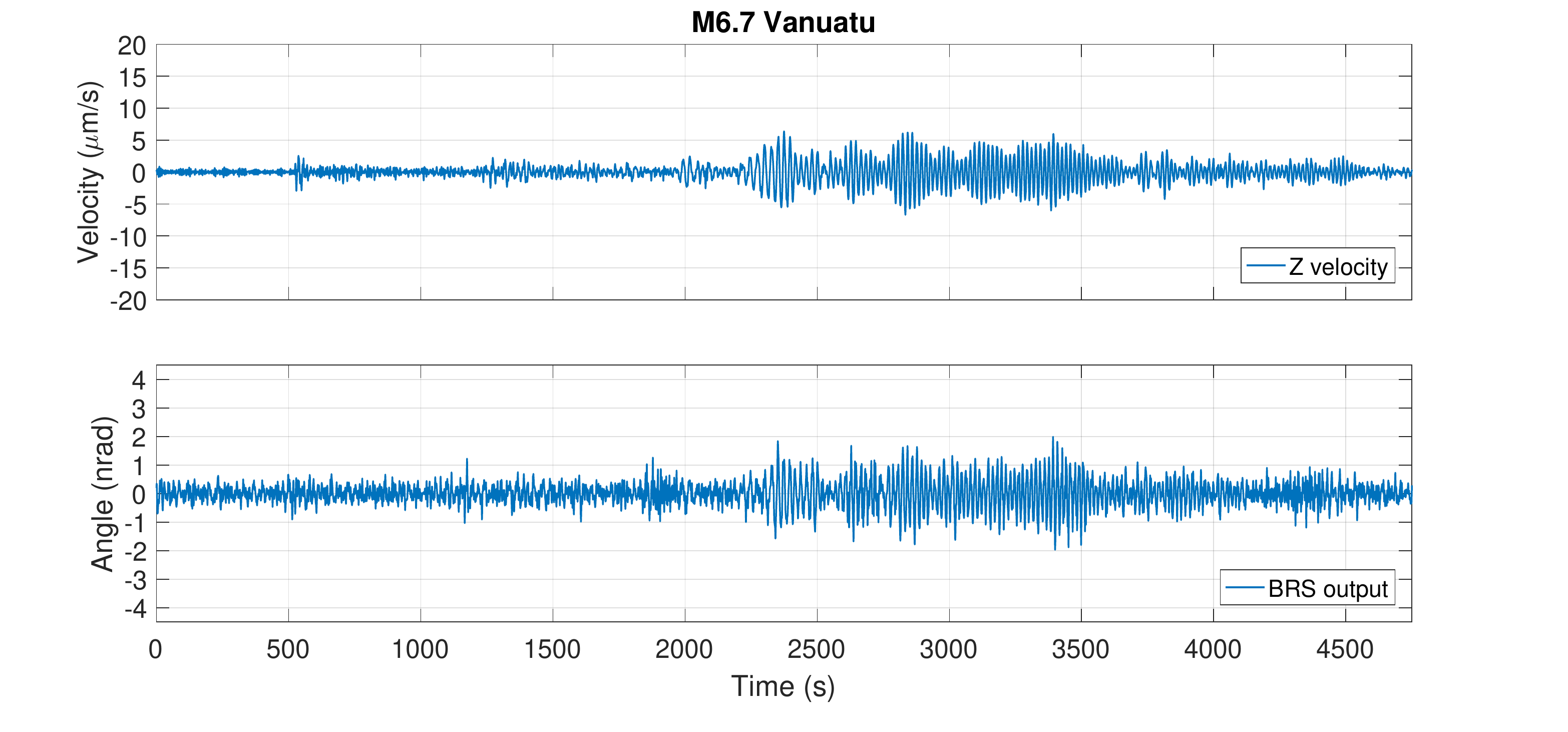}
\caption{Time series of the April 6, 2016 magnitude 6.7 earthquake 102km WSW of Sola, Vanuatu, band-pass filtered from 30\,mHz to 300\,mHz. The x-axis starts at the earthquake origin time of 6:58:48 UTC.}
\label{PTS}
\end{figure}

\begin {figure}[!h]
\centering
\includegraphics[width=1.1\textwidth]{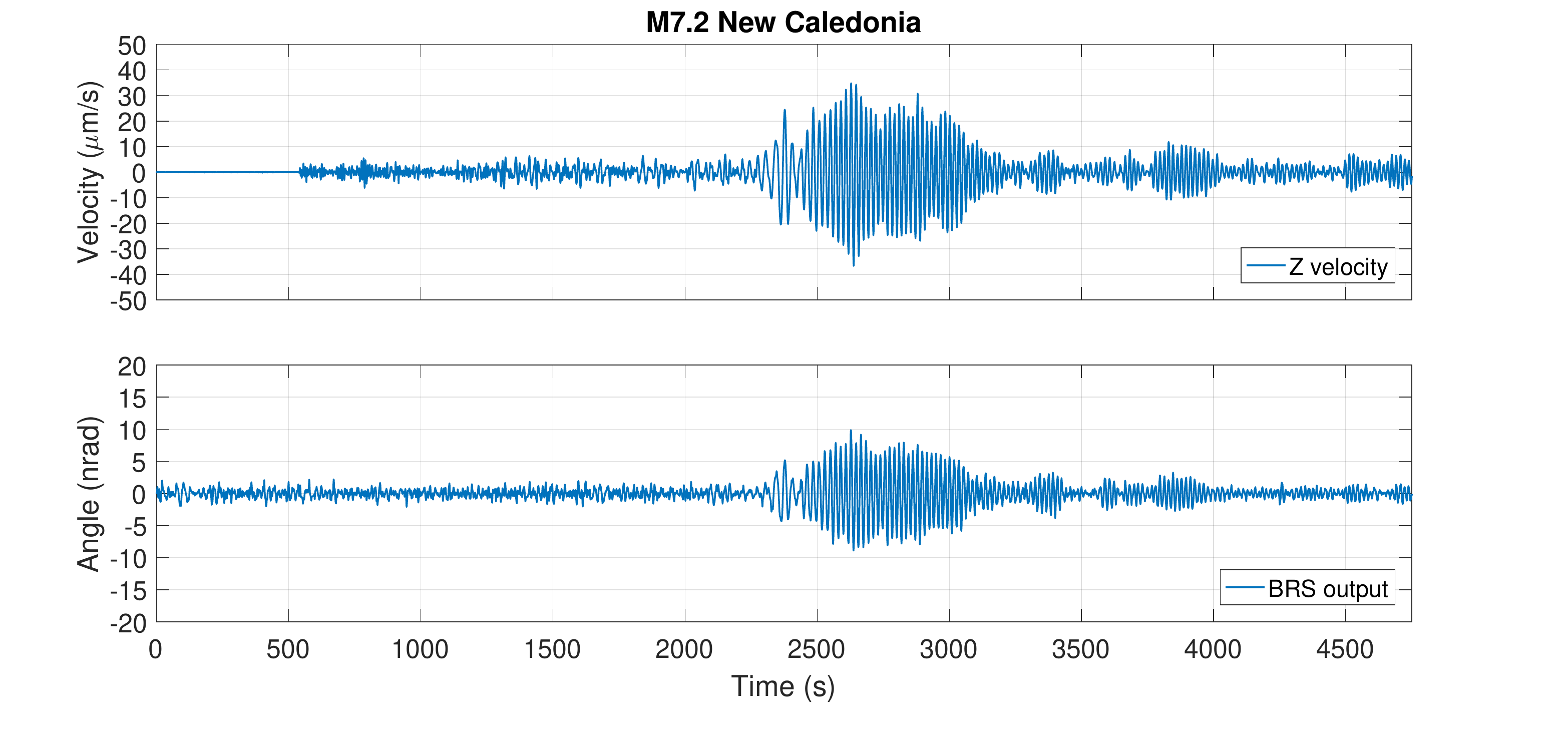}
\caption{Time series of the August 12, 2016 magnitude 7.2 earthquake 110km E of Ile Hunter, New Caledonia, band-pass filtered from 30\,mHz to 300\,mHz. The x-axis starts at the earthquake origin time of 01:26:36 UTC.}
\label{NCTS}
\end{figure}

\begin {figure}[!h]
\centering
\includegraphics[width=1.1\textwidth]{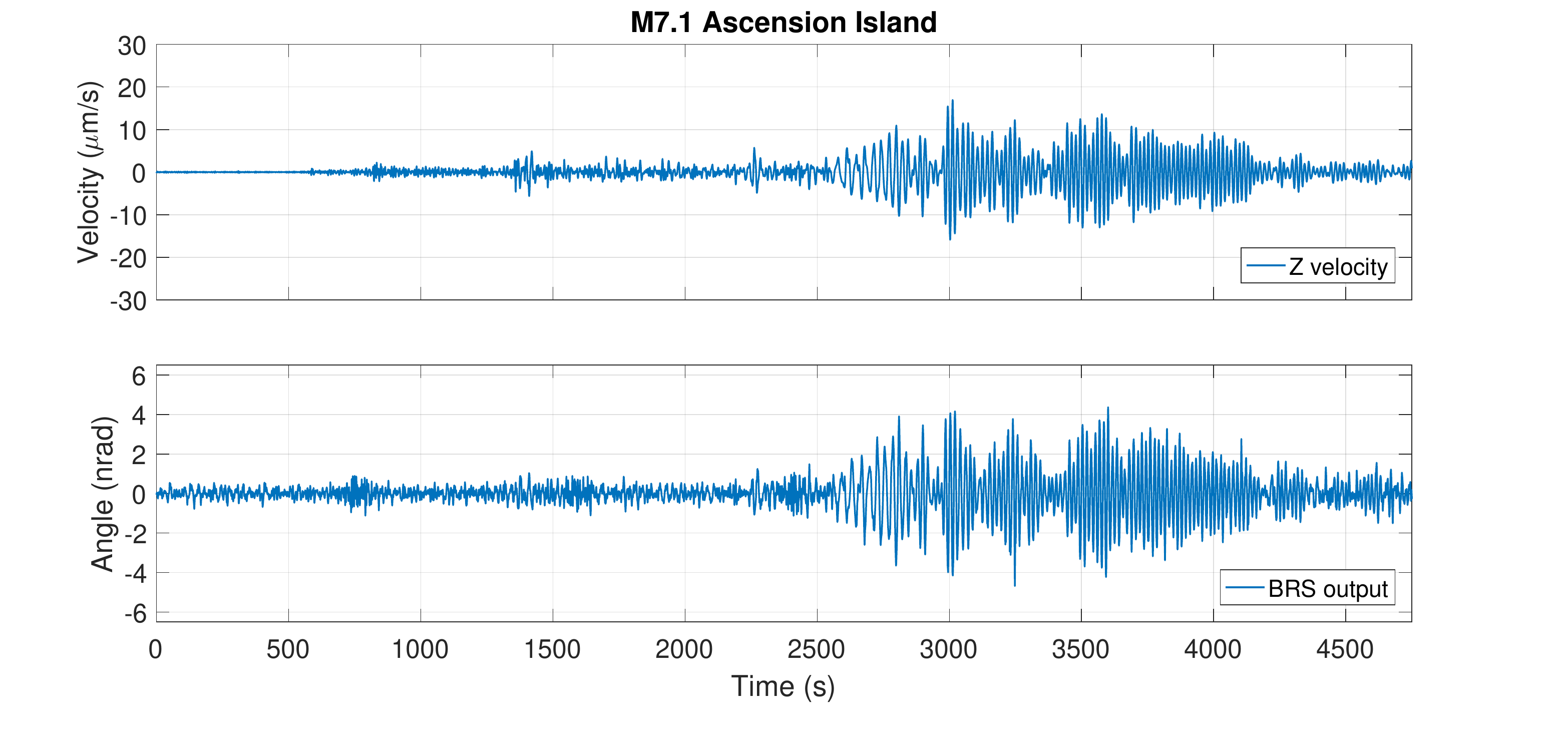}
\caption{Time series of the August 29, 2016 magnitude 7.1 earthquake North of Ascension Island, band-pass filtered from 30\,mHz to 300\,mHz. The x-axis starts at the earthquake origin time of 04:29:57 UTC.}
\label{ATS}
\end{figure}

\begin {figure}[!h]
\centering
\includegraphics[width=1.1\textwidth]{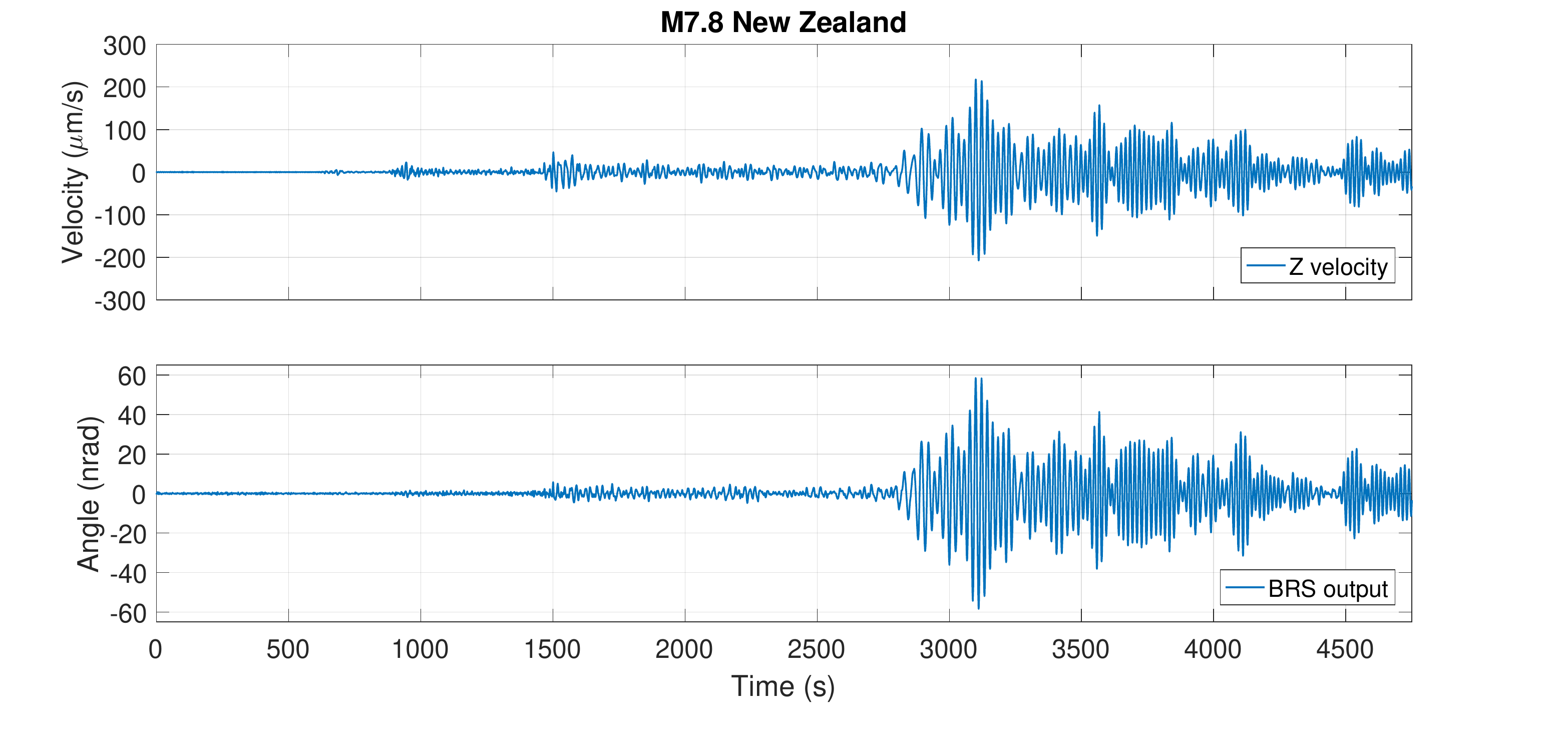}
\caption{Time series of the November 13, 2016 magnitude 7.8 earthquake 54km NNE of Amberley, New Zealand, band-pass filtered from 30\,mHz to 300\,mHz. The x-axis starts at the earthquake origin time of 11:02:56 UTC.}
\label{NZTS}
\end{figure}

\begin {figure}[!h]
\centering
\includegraphics[width=1.1\textwidth]{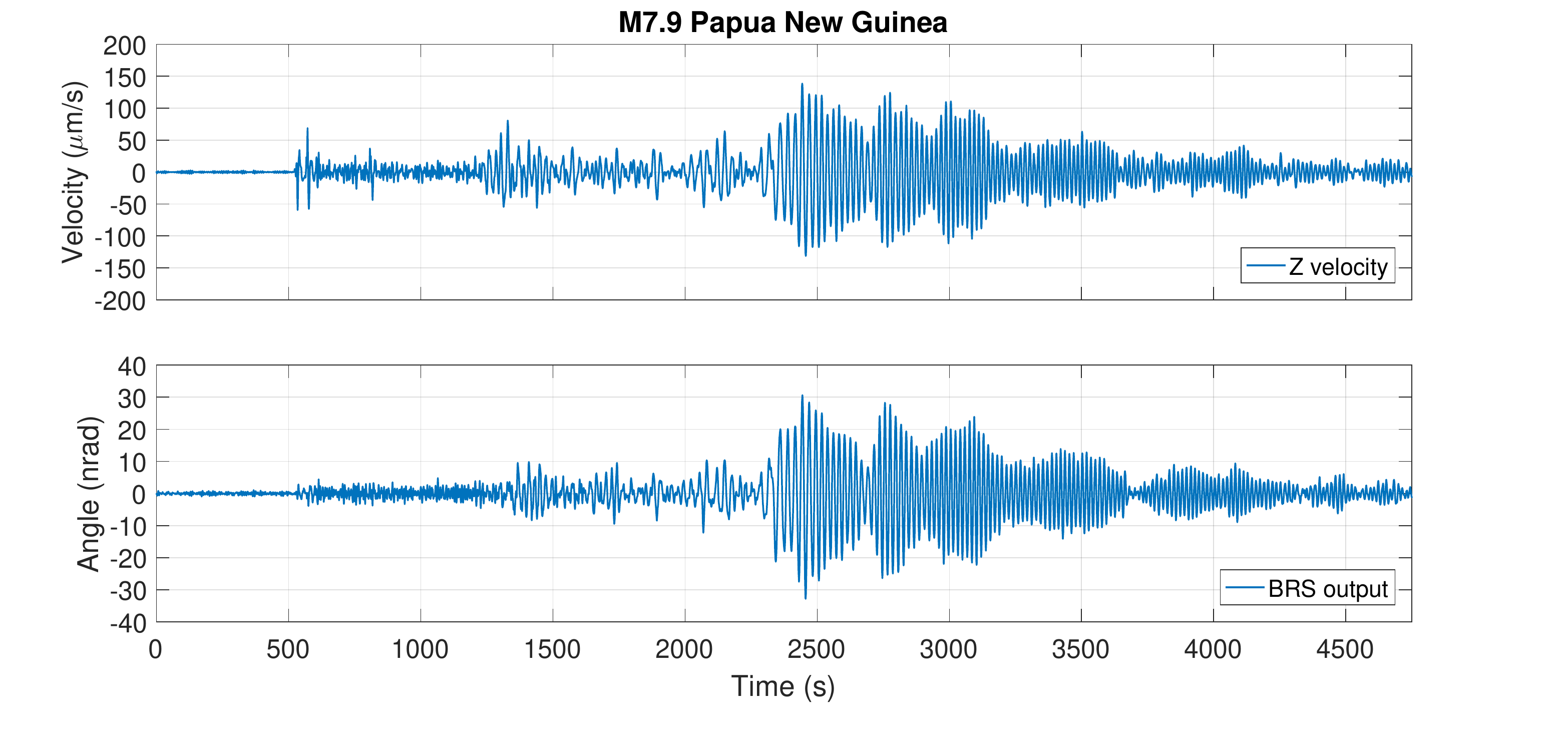}
\caption{Time series of the January 22, 2017 magnitude 7.9 earthquake 35km WNW of Panguna, Papua New Guinea, band-pass filtered from 30\,mHz to 300\,mHz. The x-axis starts at the earthquake origin time of  04:30:22 UTC.}
\label{PNG2TS}
\end{figure}

In addition to the Papua New Guinea earthquake, we have observed five other quality events shown in Figures \ref{PTS}, \ref{NCTS}, \ref{ATS}, \ref{NZTS}, and \ref{PNG2TS}. The collection of all six are those used in the analysis described in Section \ref{Analysis}. The angles of incidence listed in Table \ref{table} were estimated using Eq. \ref{SEI A} with the time delays measured by the seismometer array. All six events show almost identical signals in the Z-velocity and X-tilt.

\begin{table}[!h]
\begin{tabular}{ l l l l l}
\hline
Time of Event & Location of Epicenter & Magnitude & Measured Angle & Calculated Angle\\ 
\hline \\
2016-04-06 & 102km WSW of& 6.7 & 70.1 $^{\circ}\pm $ 2.7 $^{\circ}$ & 75.4 $^{\circ}$\\ 
06:58:48 & Sola, Vanuatu & &  & \\
\\
2016-08-12  & 110km E of  & 7.2 & 86.0 $^{\circ}\pm $ 2.0 $^{\circ}$ & 85.5 $^{\circ}$\\
01:26:36 &  Ile Hunter, New Caledonia &  &  & \\
\\
2016-08-29 & North of & 7.1&  -110.9 $^{\circ}\pm $ 0.8 $^{\circ}$ & -117.4 $^{\circ}$\\
04:29:57 & Ascension Island & &   & \\
\\
2016-11-13 & 54km NNE of & 7.8 & 89.7 $^{\circ}\pm$ 1.9 $^{\circ}$ & 98.9 $^{\circ}$\\
11:02:56 & Amberley, New Zealand & &  & \\
\\
2016-12-17  & 46km E of  & 7.9 & 67.0 $^{\circ}\pm $ 2.0 $^{\circ}$ & 59.4 $^{\circ}$\\
10:51:12 & Taron, Papua New Guinea &  & & \\

\\
2017-01-22 & 35km WNW of  & 7.9 & 56.9 $^{\circ}\pm $ 1.0 $^{\circ}$ & 61.8 $^{\circ}$\\
04:30:22 &Panguna, Papua New Guinea &  &  & \\
\\
\end{tabular}
\caption{Earthquakes captured in this analysis. Angles referenced counter clockwise from the X-axis. Measured angles are estimated using Eq. \ref{SEI A} averaged over time bins as described in Section \ref{TAnalysis} with errors representing one standard deviation.  Calculated angles are from great circle calculations. Times (UTC), magnitude, and location are courtesy of USGS.}
\label{table}
\end{table}

\clearpage

\clearpage

\pagebreak
\section{Phase Velocity Analysis}\label{Analysis}
\subsection{Temporal Analysis}\label{TAnalysis}
\quad One application of these measurements is the estimation of low frequency Rayleigh wave phase velocities. Phase velocity measurements  have direct applications in seismic wave inversion problems \cite{haskell} and are traditionally done with large arrays of translational seismometers \cite{lin}. 

Rayleigh wave phase velocities can be obtained with the measured signals by exploiting Eq. \ref{BRS V} and Eq. \ref{SEI V}. With the assumption that the incoming wave can be approximated by a single plane wave, one would expect consistent results from both equations. This allows us to provide useful local geophysical parameters while also demonstrating an application of teleseismic tilt measurements.

To extract phase dispersion curve estimates, each channel was separated into frequency bins by band-passing the time series with 5\,mHz-wide 3rd-order Butterworth filters that were centered at frequencies stepped through the band of interest (25\,mHz to 65\,mHz) in 5\,mHz steps.

\begin {figure}[!h]
\centering
\includegraphics[width=0.95\textwidth]{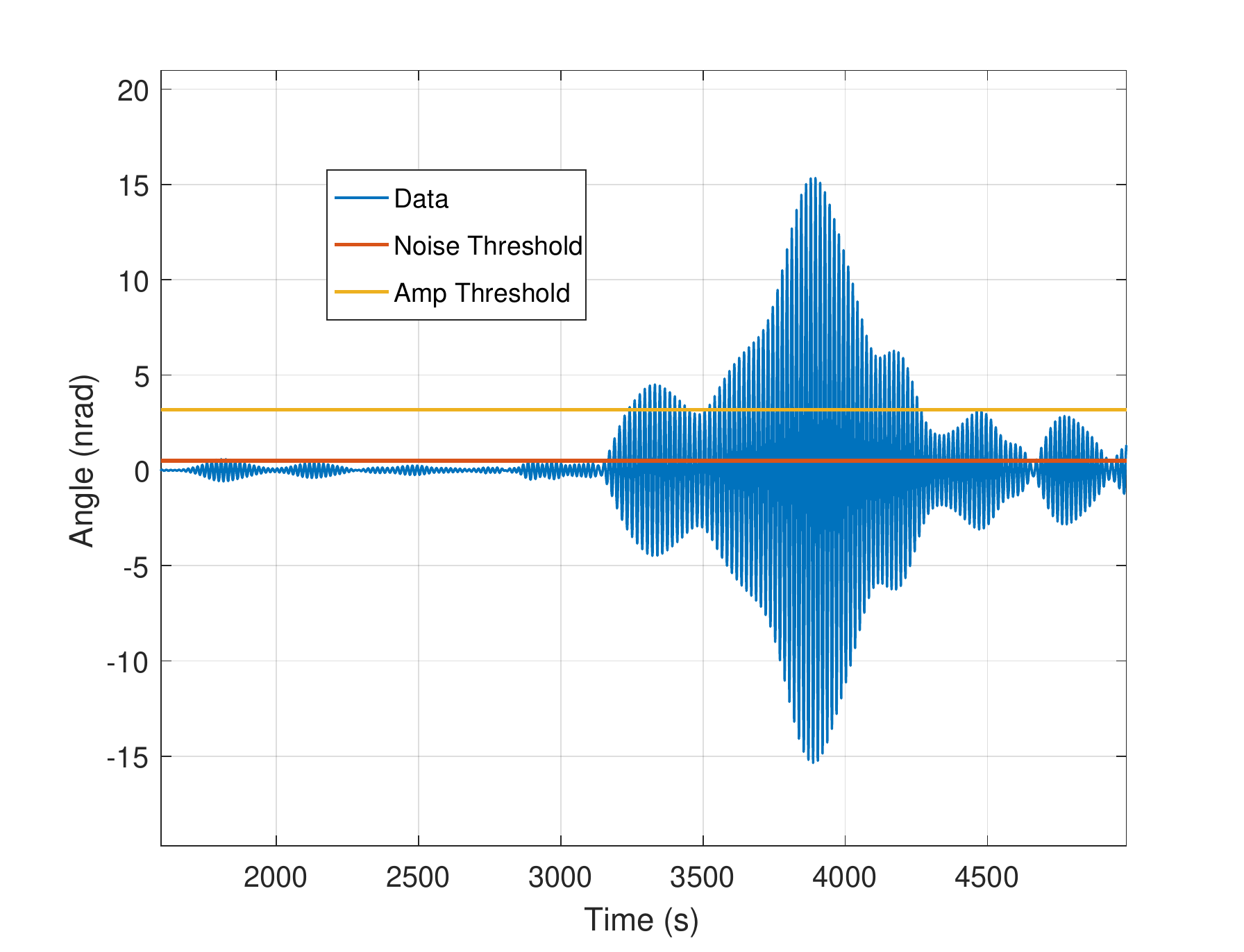}
\label{fit}
\caption{BRS data from the December 17, 2016 Papua New Guinea earthquake band-pass filtered centered at 60\,mHz along with the noise threshold, and amplitude threshold. The noise threshold is set to five times the BRS instrumental noise and is held constant for all frequencies and events. The amplitude threshold is set to 10\% of the maximum value attained by the BRS before filtering. This changes from event to event but is the same for all frequencies which focuses the analysis on the loudest sections of data as they are likely the primary shortest-path surface waves.}
\end{figure}

Phase velocity and angle of incidence estimations were first done using the array of translational seismometers. By Eq. \ref{SEI V} and \ref{SEI A}, one only needs the delays of time of arrival between the stations to reconstruct the velocity and angle since the distance is well known. These are computed by finding the center of a Gaussian that is fit to the central peak of the cross correlation of ten period-long cuts of the corner station and the end station signals for both the X-direction and Y-direction. The mean of the velocities and angles were then taken to yield a single value per frequency bin.

A second set of velocity estimations were made using the amplitudes of the tilt signal from the BRS and the vertical component of the seismometer by Eq. \ref{BRS V}. Although it is theoretically possible to obtain single-station measurements of the angle of incidence from the ratio of either the horizontal translation or rotation signals, Love-wave contamination and the lack of an orthogonal ($\theta_y$) tilt measurement required our analysis to use the angle of incidence calculated from the seismometer array. The amplitudes were computed by fitting quarter period long time series to a sinusoidal function. The resultant velocities were then averaged together for each frequency bin. 

\begin {figure}[!h]
\centering
\includegraphics[width=.85\textwidth]{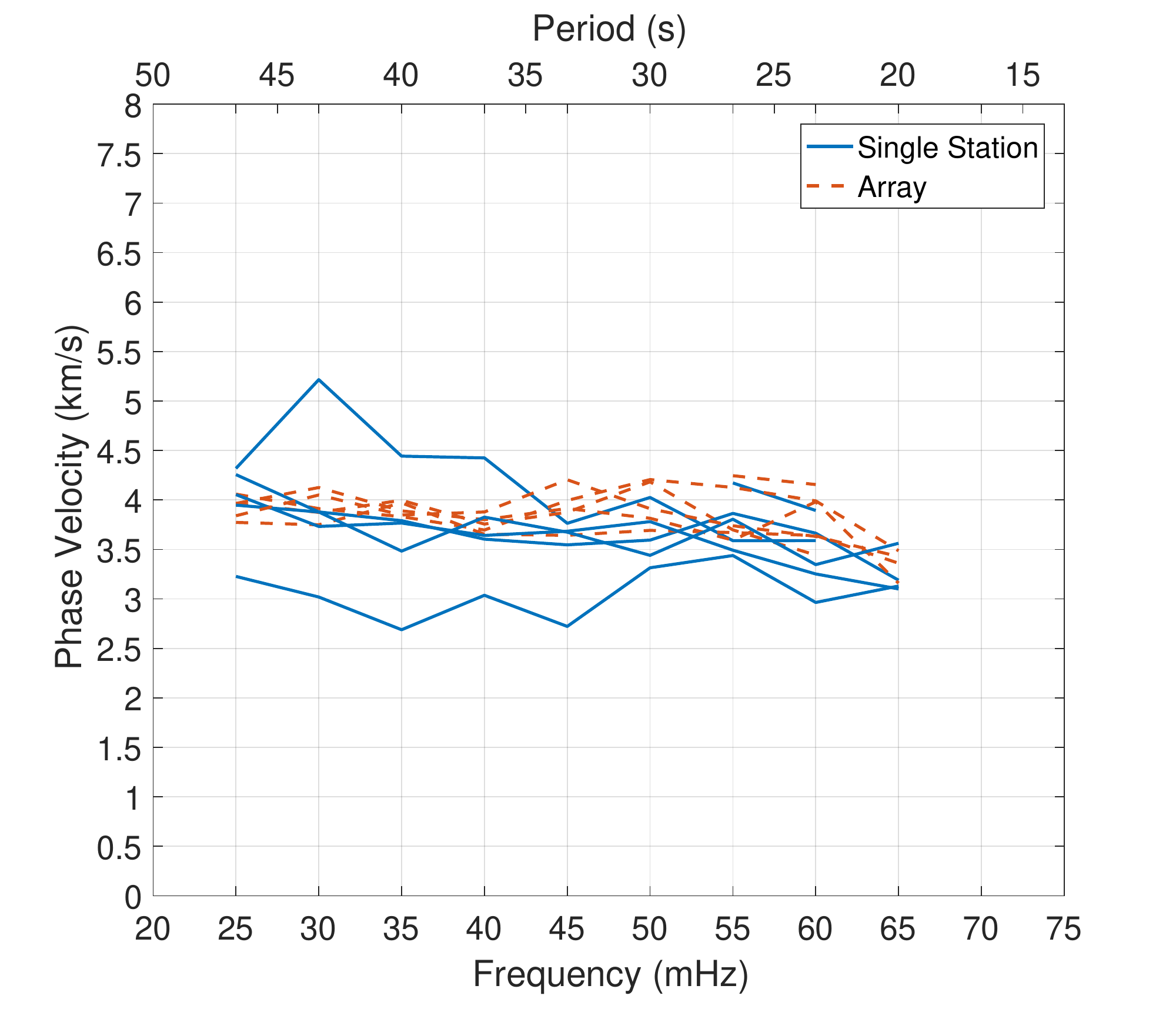}
\caption{Phase velocity for a collection of six teleseismic events calculated using the temporal analysis described in Section 5.1. ``Single station'' indicates that velocities were calculated from the ratio of vertical motion and tilt motion while ``array'' indicates velocities measured using the time delay between vertical seismometers. }
\label{Vel}
\end{figure}

Two amplitude thresholds were applied to remove sections of data that were noise dominated and those that did not contain the primary surface wave, one at 5 times the BRS sensor noise and one at 10\% of the maximum amplitude of the BRS signal before filtering. An additional cut which required the phase difference between the BRS and seismometer signals to be no more than 20$^{\circ}$ was implemented to restrict our analysis to only Rayleigh waves, as other wave types would not typically be in-phase with the tilt signal. Due to the nature of these earthquakes, our thresholds restrict each earthquake to the band for which it has prominent tilt signal. The resultant dispersion curves are shown in Fig. \ref{Vel}.

\subsection{Spectral Analysis}
\begin {figure}[!h]
\centering
\includegraphics[width=.85\textwidth]{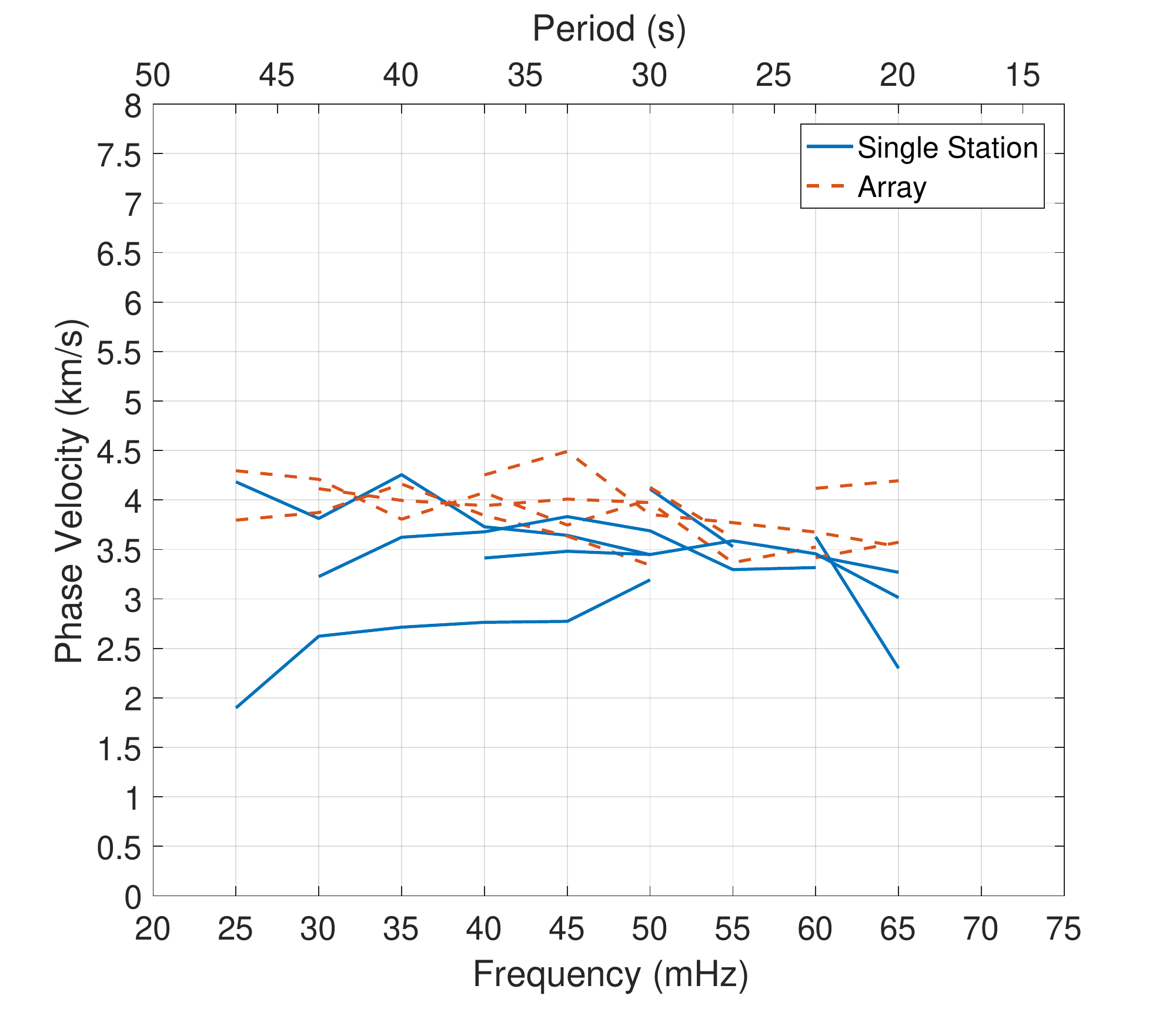}
\caption{Phase velocity for a collection of six teleseismic events calculated using the spectral analysis described in Section 5.2. ``Single station'' indicates that velocities were calculated from the ratio of vertical motion and tilt motion while ``array'' indicates velocities measured using the time delay between vertical seismometers.}
\label{VelSpec}
\end{figure}

We performed separate, more traditional, analysis which calculated phase differences and relative amplitudes from the spectral transfer function of relevant signals \cite{meier, Legendre2015}. Eqs. \ref{SEI V} and \ref{SEI A} were then used to find the corresponding phase velocities for the array and single station methods. The mean velocity in 5\,mHz wide bins were then taken to be the phase velocity for each frequency shown in Fig. \ref{VelSpec}. Bins which yielded delays with standard deviation were larger than 0.2\,s were discarded as these corresponded to bins that were noise dominated.

Although the spectral method avoids systematic errors produced by narrow filtering, it cannot distinguish between primary Rayleigh waves and other types of seismic activity. This led us to develop the temporal method described in Section \ref{TAnalysis} which, along with wave-type discrimination, can measure phase velocity evolution in time.

\section{Conclusion}
\quad We have presented high-SNR measurements of the tilt-component of surface waves from multiple teleseismic events with a ground-rotation sensor installed at LIGO Hanford Observatory and have shown that a collection of these measurements can yield consistent Rayleigh wave velocity measurements when compared to an array of seismometers. These measurements demonstrate the ability to resolve local seismological parameters using seismometers and rotation sensors at a single station and are enabled by the high tilt-sensitivity and low translational-motion coupling of the beam rotation sensor making it a useful tool in rotational seismology.

A more sophisticated analysis which takes into account the anisotropy of the earth, the multi-mode and multi-wave nature of the events, and other complex phenomena is likely to reveal further geophysical information with these datasets. Both beam rotation sensors installed at LHO are permanent pieces of the observatory's seismic isolation and will continue operation for the foreseeable future.

\section{Acknowledgments}

\quad This work was carried out at the LIGO Hanford Observatory by members of LIGO laboratory, and the LIGO Scientific Collaboration including University of Washington, Seattle, and Harvard University. LIGO was constructed by the California Institute of Technology and Massachusetts Institute of Technology with funding from the National Science Foundation (NSF), and operates under Cooperative Agreement PHY-0757058. Advanced LIGO was built under Award PHY-0823459. This document has been assigned LIGO Laboratory document number LIGO-P1700149. Participation from the University of Washington, Seattle, was supported by funding from the National Science Foundation (NSF) under Award PHY-1306613 and PHY-1607385. 

The authors would like to thank the LIGO Hanford Observatory staff for their assistance with the installation and maintenance of the sensors. We are also grateful for the useful comments and suggestions by Jan Harms and for the many thought provoking discussions with our colleagues in the E{\"o}t-Wash group. Additionally, we thank the Center for Experimental Nuclear Physics and Astrophysics (CENPA) for use of its facilities and Brian Lantz for the useful Matlab scripts. The data and analysis used in this paper can be found at: https://dcc.ligo.org/LIGO-P1700149/public.

\bibliographystyle{unsrt}
\bibliography{TiltSeismology}

\end{document}